\documentclass[aps,prb,showpacs,twocolumn]{revtex4}

\usepackage{amssymb}
\usepackage{bm} 
\usepackage{amsmath} 
\usepackage{graphicx}
\usepackage{epsfig}
\begin{document}

\title{Quantum quenches in the Dicke model: \\ statistics of the work done and of other observables}
\author{Francis N.~C.~Paraan}
\author{Alessandro Silva}
\email{asilva@ictp.it}
\affiliation{The Abdus Salam International Centre for Theoretical Physics, Strada Costiera 11, 34151 Trieste, Italy}
\date[]{Received 25 May 2009}


\begin{abstract}
We study the statistics of the work done in a zero temperature quench of the coupling constant in the Dicke model describing the interaction  between a gas of two level atoms and a single electromagnetic cavity mode. When either the final
or the initial coupling constants approach the critical coupling $\lambda_c$ that separates the normal and superradiant phases of the system, the probability distribution of the work done displays singular behavior. The average work tends to diverge as the initial coupling parameter is brought closer to the critical value $\lambda_c$. In contrast, for quenches ending close to criticality, the distribution of work has finite moments but displays a sequence of edge singularities.  This contrasting behavior is related to the difference between the processes of compression and expansion of a particle subject to a sudden change of its confining potential. We confirm this by studying in detail the time dependent statistics of other observables, such as the  quadratures of the photons and the total occupation of the bosonic modes.
\end{abstract}

\maketitle

\section{Introduction}
The study of non-equilibrium phenomena in interacting quantum systems is one of the most challenging problems
of modern condensed matter physics because several of the conceptual tools developed to describe physical systems in equilibrium ( e.g., the partition function, mean field theory, the renormalization group) are not readily generalized to non-equilibrium conditions. In order to make some progress in our understanding of the non-equilibrium behavior of these systems, it is important to identify simple paradigms of non-equilibrium processes that may be studied both theoretically and experimentally. Recently, some progress in this direction has been made through the realization of non-equilibrium experiments with cold atomic gases loaded in optical lattices\cite{Greiner, Kinoshita, Sadler} that, to a good degree of
accuracy, are well described by simple many-body models such as the Bose-Hubbard model.\cite{Bloch}

The simplest non-equilibrium process among those presently under study is the {\it quantum quench}: an abrupt change in time of one of the system parameters from an initial value $\lambda$ to a final one $\lambda'$. In a closed system, this process corresponds to the preparation of the system in the ground state $|\Psi(\lambda)\rangle$ of an initial Hamiltonian $H[\lambda]$, which  is then allowed to evolve in time according to a final Hamiltonian $H[\lambda']$. This process is particularly interesting if some qualitative changes in the state of the system between $\lambda$ and $\lambda'$ occur. This was the case in Ref.~2, where a gas of bosonic atoms was taken abruptly across a quantum critical point from the superfluid to the Mott insulating region of the phase diagram. The observation of intriguing many-body collapse and revival cycles of the two phases in the momentum distribution function signaled the high degree of many-body coherence in the dynamics of these systems.\cite{Greiner}

Theoretically, processes assimilable to quantum quenches have already been studied a few decades ago in a series of seminal papers.\cite{Mazur} More recently, however, the experiments discussed above have inspired an impressive and rapidly growing activity on this subject.\cite{Igloi, Rigol, Silva, Roux, Barankov, Polkovnikov} Apart from the concrete possibility of testing theoretical results with experiments, the main motivation behind this interest has been the shift of focus toward a
broad class of fundamental issues. More specifically, a number of recent studies addressed the extension of the concept of universality to the out of equilibrium behavior of quantum critical systems subject to either quenches at or close to criticality\cite{Igloi} or to linear sweeps of the control parameter across the quantum critical point.\cite{Zurek} Similarly, a great deal of activity is devoted to the search for dynamical manifestations of quantum integrability and to the study of the relation between non-integrability and thermalization.\cite{Rigol}

Looking for a simple and fundamental way to characterize quantum quenches from the point of view of non-equilibrium physics, it was recently observed that  a quantum quench may be considered in the context of basic statistical mechanics as a simple thermodynamic transformation.\cite{Silva,Roux,Polkovnikov} It is thus quite natural to characterize quantum quenches using standard thermodynamic variables: the work $W$ done on the system,\cite{Silva, Roux} the entropy $S$ produced,\cite{Barankov} and the heat $Q$ generated.\cite{Polkovnikov} Setting our attention on the work done, we note that since a quantum quench is not a quasistatic transformation, measurements of the work will display fluctuations among different realizations of the same process.Therefore, for a complete characterization of the quench it is necessary to specify the full probability distribution of the work done $P(W)$, which has to satisfy a number of constraints, such as the Jarzynski equalities and the Tasaki-Crooks fluctuation theorem.\cite{Jarzynski} For quantum quenches the characteristic function $G(u)=\int dW e^{-iWu} P(W)$ of the distribution $P(W)$ was found to be related to the Loschmidt echo,\cite{Silva, Talkner}
\begin{equation}
G(u) = \langle\Psi(\lambda)|\, e^{iH[\lambda] u}e^{-iH[\lambda'] u}\, |\Psi(\lambda)\rangle,
\end{equation}
a quantity that emerged previously in the study of X-ray singularities in metals,\cite{Schotte} dephasing,\cite{Karkuszewski} and quantum chaotic behavior.\cite{Peres} The Loschmidt echo is a simple and interesting object: up to a Wick rotation it can be seen as the analog of a partition function.\cite{Silva} In addition, its direct computation for a prototype quantum critical system, the Quantum Ising chain, revealed that for global quantum quenches of the transverse field the presence of criticality leads to singularities of the moments of $P(W)$ as a function of the quench parameters, while for local quenches $P(W)$ itself displays an edge singularity at low energies.\cite{Silva}

The goal of the present paper is to move one step forward toward elucidating and eventually establishing the connection between the qualitative features of the statistics of the work $P(W)$ and the generic characteristics of a physical system (e.g., its integrability, the presence of a critical point in parameter space). In order to do so, it is important to obtain benchmark results for $P(W)$ in exactly solvable models, understand its main qualitative features, and describe their physical origins. With this motivation we study the statistics of the work done in quantum quenches in the Dicke model,\cite{Dicke, Emary} an exactly solvable hamiltonian describing a gas of two level atoms interacting with a single electromagnetic cavity mode. The Dicke model is an infinite dimensional quantum critical system displaying as a function of the atom-photon
coupling a quantum phase transition between a normal and a superradiant phase.\cite{Emary, Sachdev} 

Focusing on quenches of the coupling constant within the normal phase, we show that criticality leaves clear signatures in the dynamics of the system and on $P(W)$: as the initial coupling tends to the critical point the average work done on the system tends to diverge, while for quenches ending at criticality the probability distribution displays an interesting sequence of edge singularities. We will give a simple physical picture explaining this difference, ultimately related to the difference between the processes of compression and expansion of a particle subject to a sudden change of its confining potential. We further elucidate these findings by computing exactly the time-dependent statistics of observables such as the quadrature operators of the photon field and the occupation of the bosonic  modes. The rest of the paper is organized as follows: in Sec.~\ref{sec1} we introduce the model, establish the notations, and discuss the statistics of the work $P(W)$. In Sec.~\ref{sec2} we study the statistics of the observables starting with the quadratures of the cavity field followed by the total occupation of the bosons. Finally, in Sec.~\ref{sec3} we give our conclusions.

\section{The Hamiltonian and the statistics of the work done}\label{sec1}

The Dicke model\cite{Dicke} describes the coupling of $N$ two level systems, such as two level atoms, to a single electromagnetic cavity mode.  Its hamiltonian under the condition of resonant coupling, which simplifies the succeeding analysis, is given by
\begin{equation}\label{hamiltonian}
H = \sum_{i=1}^N\omega_0\sigma_i^z + \omega_0a^\dagger a + \frac{\lambda}{\sqrt{N}}\sum_{i=1}^N(a^\dagger + a)(\sigma_i^+ + \sigma_i^-),
\end{equation}
where the Pauli matrices $\sigma_i$ describe the dynamics of the two level systems, while $a$ ($a^\dagger$) destroys (creates) a photon of frequency $\omega_0$. The coupling between atoms and photons has been rescaled by $1/\sqrt{N}$ in order to have a  well-defined thermodynamic limit $N\to +\infty$. Here and throughout this work a natural system of units is employed in which action is measured in units of $\hbar$.

The Dicke model has two phases,\cite{Dicke, Emary} a normal and a superradiant one that are separated by a quantum critical point at $\lambda_c = \omega_0/2$. The transition as $\lambda$ exceeds $\lambda_c$ is characterized by the breaking of parity symmetry leading to the spontaneous generation of an extensive density of photons in the system $\langle a^\dagger a \rangle \propto N$.

The Hamiltonian Eq.~(\ref{hamiltonian}) can be easily diagonalized exactly.\cite{Dicke}  Focusing on the normal phase ($\lambda < \lambda_c$) in the thermodynamic limit,\cite{Emary} it is first of all convenient to regroup the Pauli matrices into collective spin operators $J^q = \sum_{i=1}^N \sigma_i^q$, where $q=z,\pm$. Using now the Holstein-Primakoff representation in terms of a bosonic mode $b$, $J^z = b^\dagger b-N/2$, $J^+ = b^\dagger\sqrt{N-b^\dagger b}$, $J^-= \sqrt{N-b^\dagger b}\; b$, the semiclassical/thermodynamic limit $N\to +\infty$ can be taken. One obtains
\begin{equation}
H = \omega_0(a^\dagger a + b^\dagger b) + \lambda(a^\dagger +a)(b^\dagger +b) - \frac{N\omega_0}{2}.
\end{equation}
At this point, the diagonalization proceeds by a standard Bogoliubov rotation. The final form of the Hamiltonian is
\begin{equation}
H[\lambda] = \omega_+c_+^\dagger c_+ + \omega_-c_-^\dagger c_- + C.
\end{equation}
Here the eigenenergies are given by
\begin{equation}
\omega_{\pm}(\lambda) = \omega_0\sqrt{1\pm\frac{2\lambda}{\omega_0}},
\end{equation}
while the eigenmodes $c_\pm$ can be expressed in terms of $a$ and $b$ as
\begin{equation}\label{eigenmodes}
c_\pm = \cosh(r_\pm) \frac{a\pm b}{\sqrt{2}} + \sinh(r_\pm)\frac{a^\dagger \pm b^\dagger}{\sqrt{2}},
\end{equation}
with $\tanh(r_\pm) = (\omega_\pm - \omega_0)/(\omega_\pm + \omega_0)$. Finally, the constant is $C = [\omega_+ + \omega_- -\omega_0(N +2)]/2$. The vanishing of $\omega_-$ at $\lambda_c = \omega_0/2$ is a direct consequence of the presence of a quantum critical point. Though we will not consider the superradiant phase at length here, we note that for $\lambda > \lambda_c$ a consistent thermodynamic limit can be taken only after the bosonic operators $a,b$ are displaced.\cite{Emary}

In the following analysis we will be interested in quantum quenches of the coupling constant from an initial value $\lambda$
to a final one $\lambda'$. We will focus on the case where both  $\lambda,\lambda' < \lambda_c$. From the point of view of the statistics of the work done, quantum quenches from $\lambda<\lambda_c$ towards $\lambda' >\lambda_c$  (or vice versa) are not very interesting because the generation of a photon density $\propto N$ requires a work that scales extensively with the number of atoms, while fluctuations are expected to scale like $1/\sqrt{N}$, i.e., to be highly suppressed in the thermodynamic limit. On the other hand, we do not expect major changes in the main qualitative results
of this paper for quenches with both $\lambda,\lambda' > \lambda_c$: the choice of focusing on $\lambda,\lambda' <\lambda_c$ has the only purpose of allowing us to obtain closed analytic results for the statistics of the work and of other observables.

\begin{figure}
	\centering
		\includegraphics[width=1\linewidth]{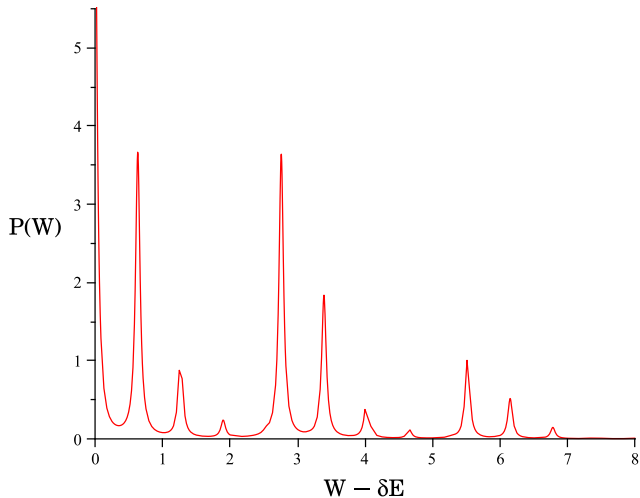}
		\caption{A typical plot of the probability distribution of the work $P(W)$ for a quench from $\lambda=0.5$ to $\lambda' =0.9$ (the delta functions have been Lorentz-broadened for clarity and we have set $\omega_0 = 1$). Notice the sequence of principal peaks separated by $2\omega_+(\lambda') \simeq 2.75$, each followed by a tail of subpeaks with separation $2\omega_-(\lambda') \simeq 0.63$. When $\lambda' = \lambda_c$ these subpeaks merge and give rise to an edge singularity at each principal peak (see Fig.~\ref{edge}).}
	\label{distribution}
\end{figure}

Let us start our analysis by computing and characterizing qualitatively the statistics of the work done in a quantum quench of the coupling constant from $\lambda$ to $\lambda'$ $(\lambda,\lambda'<\lambda_c)$. As stated earlier in the introduction, in order to study the probability distribution $P(W)$ of the work $W$ done in a quantum quench it is convenient to compute its
characteristic function $G(u)=\int dW e^{-i W u} P(W) $. In terms of the initial and final Hamiltonians $H[\lambda]$ and $H[\lambda']$, the latter is given by the Loschmidt echo,\cite{Silva} 
\begin{equation}
G(u)  = \langle 0_\lambda|\, e^{iH[\lambda]u} e^{-iH[\lambda']u}\,|0_\lambda\rangle,
\end{equation}
where $|0_\lambda\rangle$ is the vacuum of the operators $c_\pm$. For $\lambda,\lambda' <\lambda_c$ the operators diagonalizing the final hamiltonian $\bar{c}_\pm$ are related to the initial eigenmodes $c_\pm$ by a Bogoliubov rotation
\begin{equation}\label{relation}
\bar{c}_\pm = \cosh(\xi_\pm) c_\pm + \sinh(\xi_\pm)c_\pm^\dagger,
\end{equation}
with $\tanh(\xi_\pm) = {[}\omega_\pm(\lambda') - \omega_\pm(\lambda){]}/{[}\omega_\pm(\lambda') + \omega_\pm(\lambda){]}$.

From these definitions it is evident that we have to compute
\begin{equation}
G(u) = e^{-i\delta E u}\langle0_\lambda| e^{-i[\omega_+(\lambda')\bar{c}_+^\dagger \bar{c}_+ + \omega_-(\lambda')\bar{c}_-^\dagger \bar{c}_-]u}|0_\lambda\rangle,
\end{equation}
where $\delta E$ is the difference in the ground state energies of the initial and final hamiltonians. In order to do so one has first to express $|0_\lambda\rangle$ in terms of the vacuum $|0_{\lambda'}\rangle$ of the final eigenmodes $\bar{c}_\pm$. Using the definition ${c}_\pm |0_{\lambda}\rangle = 0$ together with the Bogoliubov rotation Eq.~(\ref{relation}), one obtains the equation $\cosh(\xi_\pm)\bar{c}_\pm|0_{\lambda}\rangle = \sinh(\xi_\pm)\bar{c}_\pm^\dagger|0_{\lambda}\rangle $, which implies that
\begin{equation}\label{vacuum}
|0_{\lambda}\rangle = S_+[\xi_+]S_-[\xi_-]|0_{\lambda'}\rangle,
\end{equation}
where 
\begin{equation}
S_\pm[z] = e^{\frac{1}{2}[zc_\pm^\dagger(\lambda')^2 - z^*c_\pm(\lambda')^2]},
\end{equation}
are single-mode squeezing operators.\cite{Barnett} The state $|0_\lambda\rangle$ is therefore a squeezed vacuum of the modes $\bar{c}_\pm$. In this representation the Loschmidt echo takes the simple form
\begin{equation}
G(u) = \langle 0_{\lambda'}| S_+^\dagger[\xi_+]S_-^\dagger[\xi_-]S_+[\xi_+(u)]S_-[\xi_-(u)] |0_{\lambda'}\rangle, 
\end{equation}
where $\xi_\pm(u) = \xi_\pm e^{-2i\omega_\pm(\lambda')u}$. Using standard formulas for the overlap of squeezed states we then obtain
\begin{equation}\label{char2}
G(u) = e^{-i\delta Eu}G_+(u)G_-(u),
\end{equation}
where
\begin{equation}\label{charpm}
G_{\pm}(u) = \bigl[1+ \bar{n}_\pm - e^{-2i\omega_{\pm}(\lambda')u} \bar{n}_\pm \bigr]^{-\frac{1}{2}}.
\end{equation}
Here we introduced the parameters
\begin{equation}
\bar{n}_\pm = \sinh^2(\xi_\pm) = \frac{[\omega_\pm(\lambda') - \omega_\pm(\lambda)]^2}{4\omega_\pm(\lambda')\omega_\pm(\lambda)},
\end{equation}
which physically represent the average occupation of the final eigenmodes $\bar{c}_\pm$ in the initial ground state $| 0_\lambda\rangle$.

From these equations one can immediately deduce that the distribution $P(W)$ has the form
\begin{multline}
P(W) = \sum_{k,l=0}^{+\infty} \mathcal{P}_+(2k)\mathcal{P}_-(2l)\\\times\delta[W-\delta E-2k\omega_+(\lambda') - 2l\omega_-(\lambda')],
\end{multline}
where
\begin{equation}
\mathcal{P}_\pm (2k)= \frac{1}{\sqrt{1+\bar{n}_\pm}} \binom{k-\tfrac{1}{2}}{k}\biggl[\frac{\bar{n}_\pm}{1+\bar{n}_\pm}\biggr]^k. 
\end{equation}
Qualitatively,  the distribution $P(W)$ consists of a series of principal peaks separated by $2\omega_+(\lambda')$, each followed by a tail of subpeaks describing excited $-$ modes that are separated by $2\omega_-(\lambda')$ (see Fig.~\ref{distribution}).

The partial amplitudes $\mathcal{P}_\pm$ control the weight of each peak in $P(W)$. The presence of a quantum critical point
and its effect on $P(W)$ can be elucidated by studying the asymptotic behavior of $\mathcal{P}_\pm(k)$ for large $k$. Using Stirling's formula, $z!\approx \sqrt{2\pi}z^{z+1/2}e^{-z}$ we obtain
\begin{equation}
\binom{k-\tfrac{1}{2}}{k} = \frac{(2k)!}{2^{2k}(k!)^2} \approx \frac{1}{\sqrt{\pi k}}, 
\end{equation}
from which, for $k \gg 1$, one gets
\begin{align}
	\mathcal{P}_\pm(2k) &\approx \frac{1}{\sqrt{1+\bar{n}_\pm}}\frac{e^{-k/\zeta_\pm}}{\sqrt{\pi k}}, \label{powerlaw} \\
	\zeta_\pm^{-1} &= \log\biggl[1 +\frac{1}{\bar{n}_\pm} \biggr].
\end{align}
The scale $\zeta_\pm$ controls the decay of the corresponding amplitude. Notice now that the vanishing of $\omega_-(\lambda)$ ${\bm(} \omega_-(\lambda'){\bm)}$ at the critical coupling implies the divergence of $\bar{n}_-$ when $\lambda \to\lambda_c$ ${\bm(}\lambda' \to\lambda_c {\bm)}$. Therefore when $\lambda\to\lambda_c$ we have
\begin{equation}
\zeta_- \approx \bar{n}_- \approx \frac{\omega_-(\lambda')}{4\omega_-(\lambda)} \propto \sqrt{\frac{\lambda_c}{\lambda_c-\lambda}},
\end{equation}
and a similar equation with $\lambda' \leftrightarrow \lambda$ for $\lambda'\to\lambda_c$. The presence of a quantum phase transition in parameter space is marked by the divergence of the scale controlling the exponential decay of the  partial amplitude associated with the $-$ modes, which are the the ones becoming critical at the transition. Notice that when either $\lambda$ or $\lambda'$  is exactly at the quantum critical point the partial amplitude $\mathcal{P}_-$ decays as a power law,  as seen in Eq.~(\ref{powerlaw}).

\begin{figure}[!t]
	\centering
		\includegraphics[width =1\linewidth]{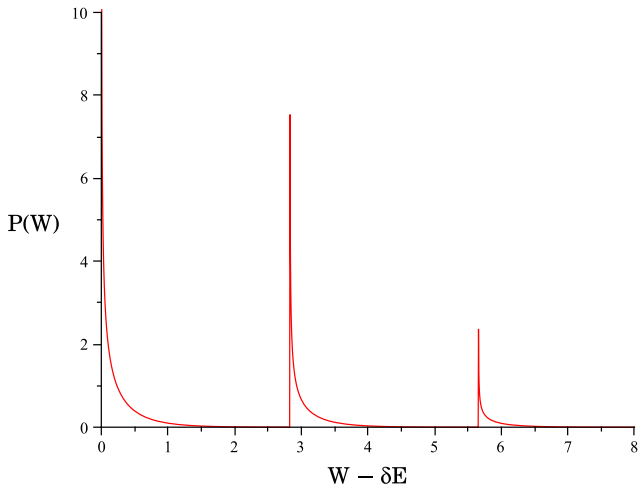}
		\caption{The probability distribution of the work $P(W)$ for a quench from $\lambda = 0.1$ to $\lambda' =\lambda_c$ (here we set $\omega_0 = 1$).  A sequence of edge singularities described by Eq.~(\ref{singular}) and separated by $2\omega_+(\lambda_c)=2.75$ signals the criticality of the system in the final state.}
	\label{edge}
\end{figure}

Despite the fact that for both quantum quenches toward the quantum critical point and away from the quantum critical point $\zeta_-$ diverges, the physics behind these two situations is deeply different. Indeed, while for quenches starting close to criticality all the cumulants of the distribution $P(W)$ diverge, quenches towards criticality are marked by the presence of a sequence of edge singularities close to each principal peak. At the root of this difference, explained in detail below, is the fact that for quenches toward criticality the divergence of $\zeta_-$ is accompanied by the vanishing of $\omega_-(\lambda')$ as $\lambda'\to\lambda_c$. Hence the energy scale $\Omega_- = 2\zeta_-\omega_-(\lambda')$ controlling the decay of the tails of $P(W)$ remains finite
\begin{equation}\label{omegamin1}
\Omega_- \simeq \frac{\omega_-(\lambda)}{4},
\end{equation}
as $\lambda'\to\lambda_c$.
On the other hand, for quenches starting infinitesimally close to criticality ($\lambda \to \lambda_c$) we have
\begin{equation}
\Omega_- \simeq \frac{[\omega_-(\lambda')]^2}{4\omega_-(\lambda)} \propto \frac{\lambda_c}{\lambda_c - \lambda}.
\end{equation}

For quenches starting at criticality ($\lambda\to \lambda_c$), the divergence of $\Omega_-$ implies the divergence of the average work done on the system. Physically, this is due to the fact that when a quench starts exactly at criticality the mode $\bar{c}_-$ corresponds in coordinate space to a completely delocalized free particle. Every quench to $\lambda' \ne\lambda_c$ describes the switching of an harmonic potential tending to confine the mode in a finite volume. This process is conceptually similar to the compression of a gas occupying an infinite volume into a finite one: on this basis we expect the average work done to diverge as
$\lambda\to\lambda_c$. In order to see this, we may use the characteristic function $G(t)$ to extract directly the cumulants
$K_n$ of the distribution $P(W)$ using the standard cumulant expansion formula $G(u) = \exp\bigl[\sum_{n=1}^{\infty}(-iu)^n/n!\,K_n \bigr]$. Expanding Eq.~(\ref{charpm}) to first order we obtain
\begin{equation}
\langle W\rangle = K_1 = \delta E + \omega_+(\lambda')\bar{n}_+ + \omega_-(\lambda')\bar{n}_-. 
\end{equation}
The average  excess work $\langle \delta W\rangle = \langle W \rangle -\delta E $ is then
\begin{equation}
\langle\delta W\rangle = \frac{[\omega_+(\lambda') - \omega_+(\lambda)]^2}{4\omega_+(\lambda)} + \frac{[\omega_-(\lambda') - \omega_-(\lambda)]^2}{4\omega_-(\lambda)}. 
\end{equation}
From this expression it is evident that quantum quenches are not reversible processes, $W(\lambda\to\lambda') \ne -W(\lambda'\to\lambda)$, and that for $\lambda \to\lambda_c$ one has
\begin{equation}
\langle \delta W\rangle \sim \frac{[\omega_-(\lambda')]^2}{\sqrt{2\omega_0}} \frac{1}{\sqrt{\lambda-\lambda_c}}, 
\end{equation}
which diverges at criticality as anticipated. A similar divergence is observed in the  second moment of $P(W)$, which can
be easily computed to find
\begin{equation}
\langle (\delta W)^2 \rangle = 2[\omega_+(\lambda')]^2\bar{n}_+(1+\bar{n}_+) + 2[\omega_-(\lambda')]^2\bar{n}_-(1+ \bar{n}_-). 
\end{equation}

A diametrically opposite situation is obtained for quenches starting with $\lambda \ne\lambda_c$ and ending at criticality.
In this case, the initial potential confining the mode $c_-$ is removed in the final state. This process is conceptually
similar to the expansion of a gas in vacuum and is therefore not expected to be characterized by a divergence of the average work. However, it turns out that the resulting distribution $P(W)$ displays a series of edge singularities resulting from the fact that $\Omega_-$ stays finite even in the limit $\lambda' \to\lambda_c$ (see Eq.~(\ref{omegamin1})). The most elegant way to obtain  this result is by observing that for $u\ll 1/\omega_-(\lambda')$ we can always approximate
\begin{equation}
G_-(u) \simeq (1+i\Omega_-u)^{-1/2},
\end{equation}
whose Fourier transform is
\begin{equation}\label{singular}
P_-(w) = \frac{\Theta(w)}{\sqrt{\pi w \Omega_-}}\,e^{-w/\Omega_-}. 
\end{equation}
Using these expressions for $\lambda' = \lambda_c$ one immediately obtains
\begin{equation}
P(W) = \sum_{k=0}^{+\infty}\mathcal{P_+}(2k)P_-(W-\delta E -\sqrt{8}\omega_0 k), 
\end{equation}
which has the expected form (see Fig.~\ref{edge}). It is interesting to notice that edge singularities in the statistics of the work have been previously reported in local quenches of the transverse field in a Quantum Ising chain.\cite{Silva} These two examples seem to suggest a connection between criticality and edge singularities in the statistics of the work done in quantum quenches, a topic which deserves a deeper study in the future.

\section{Statistics of other observables}\label{sec2}

Let us now continue the characterization of quantum quenches in the Dicke model by focusing on the statistics of observables such as the quadrature operators and the occupation of the bosonic modes. In contrast with the statistics of the work, which is time independent, the statistics of generic observables depends on the time $t$ elapsed after the quench. More explicitly, let us consider a generic observable $\hat{Q}$ having eigenstates $|n\rangle$ with corresponding eigenvalues $q_n$. If the initial state before the quench is $|\Phi_0\rangle$, the probability to obtain $q$ as a result of the measurement of $\hat{Q}$ at time $t$ is
\begin{equation}
P(q,t) = \sum_{n} \bigl| \langle n| e^{-iH_f t} | \Phi_0 \rangle \bigr|^2 \delta(q-q_n),
\end{equation}
where $H_f$ is the final hamiltonian. Hence, the characteristic function $G_Q(u, t) = \int e^{-iqu} P(q,t)\,dq$ of the distribution $P(q,t)$ is given by the expression
\begin{equation}
G_Q(u,t) = \langle\Phi_0| e^{-i\hat{Q}(t) u} |\Phi_0\rangle,
\end{equation}
where $\hat{Q}(t) = e^{iH_ft}\hat{Q}e^{-iH_ft}$.

Let us now compute explicitly this characteristic function for two  important observables for the Dicke model: the quadratures given by
\begin{equation}\label{quad}
X(\alpha) = \frac{1}{\sqrt{2}}(ae^{-i\alpha} + a^\dagger e^{i\alpha}),
\end{equation}
and the total occupation of the bosonic modes
\begin{equation}
N = a^\dagger a + b^\dagger b,
\end{equation}
which can be interpreted as the order parameter of the transition.
 
\subsection{Quadrature operators}
Let us start by computing the statistics of the quadrature operator $X(\alpha)$ for a generic quench starting
at $\lambda$ and ending at $\lambda'$ with $\lambda,\lambda'<\lambda_c$. The appropriate characteristic function is
\begin{equation}
G_\alpha(u,t) = \langle0_\lambda|\, e^{iH[\lambda']t}e^{-iX(\alpha) u} e^{-iH[\lambda']t}\,|0_\lambda\rangle.
\end{equation}
Since the final hamiltonian is diagonalized by the modes $\bar{c}_\pm$, it is convenient to first express the operator $X(\alpha)$ in Eq.~(\ref{quad}) in terms of them.  Inverting a transformation analogous to Eq.~(\ref{eigenmodes}) (with $\lambda \rightarrow \lambda'$) yields
\begin{multline}
	a = \frac{1}{\sqrt{2}}\bigl[\cosh(\bar{r}_+)\bar{c}_+ +\cosh(\bar{r}_-)\bar{c}_- \\ -\sinh (\bar{r}_+)\bar{c}_+^\dagger -\sinh(\bar{r}_-)\bar{c}_-^\dagger \bigr], 
\end{multline}
 where the bars mean that the corresponding quantities are to be evaluated using the final value of the coupling parameter $\lambda'$. Since the $\bar{c}_\pm$ operators evolve trivially in time according to the final Hamiltonian, we find that the characteristic function has the simple expression
\begin{equation}
G_\alpha(u,t) = \langle0_\lambda|\, e^{-iX_+(\alpha,t)u}e^{-iX_-(\alpha,t)u}\,|0_\lambda\rangle,
\end{equation}
where
\begin{equation}
X_\pm(\alpha,t) = \frac{A_\pm(\alpha)}{2}\,e^{-i\omega_\pm(\lambda')t}\bar{c}_{\pm} + \frac{A_\pm^*(\alpha)}{2}\,e^{i\omega_\pm(\lambda')t}\bar{c}_{\pm}^\dagger,
\end{equation}
with
\begin{equation}
A_\pm(\alpha) = \bigl[ e^{-i\alpha} \cosh(\bar{r}_\pm) -e^{i\alpha} \sinh(\bar{r}_\pm)\bigr]. 
\end{equation}
The next step in obtaining a closed form result consists of expressing the state $|0_\lambda\rangle$ in terms of the vacuum of the operators $\bar{c}_\pm$. Using the result of the previous section (Eq.~(\ref{vacuum})), we obtain
\begin{equation}
G_\alpha(u,t) = \langle0_{\lambda'}|\, e^{-i\tilde{X}_+(\alpha,t)u}e^{-i\tilde{X}_-(\alpha,t)u}\,|0_{\lambda'}\rangle,
\end{equation}
where $\tilde{X}(\alpha,t) = S^\dagger(\xi_\pm) X_\pm(\alpha, t)S(\xi_\pm)$. Finally, using the formula $S^\dagger(\xi)\, c\, S(\xi) = \cosh(\xi)\,c + \sinh(\xi)\,c^\dagger$, we obtain
\begin{equation}
G_\alpha(u,t) = \langle e^{\eta_+u\bar{c}_+-\eta_+^*u\bar{c}_+^\dagger}\,e^{\eta_-u\bar{c}_- -\eta_-^*u\bar{c}_-^\dagger}\rangle,
\end{equation}
with
\begin{multline}
\eta_{\pm}(\alpha) = -\frac{i}{2}\bigl[A_\pm(\alpha) e^{-i\omega_\pm t}\sinh(\xi_\pm) \\ 
+ A_\pm^*(\alpha)e^{i\omega_\pm t}\cosh(\xi_\pm)\bigr].
\end{multline}
Since the exponentials appearing in the expression Eq.~(\ref{char2}) for the characteristic function are standard displacement operators, taking their vacuum expectation value gives\cite{Barnett}
\begin{equation}
G_\alpha(u,t)  = e^{-\frac{u^2}{2}[|\eta_+(\alpha,t)|^2 + |\eta_-(\alpha,t)|^2]}.
\end{equation}
The statistics of the quadrature operators maintains the characteristics it has in the initial state and is always
gaussian. The only scale characterizing the distribution is its variance, which is given by
\begin{equation}
\langle\delta X(\alpha)^2\rangle = \bigl|\eta_+(\alpha,t)\bigr|^2 + \bigl|\eta_-(\alpha,t)\bigr|^2
\end{equation}

Using this expression we may get further insight on the difference between  quenches that are driven toward criticality and those that start near criticality and then are driven away from it. Indeed, focusing on the case $\alpha =0$ that corresponds to the ``coordinate'' operator $X(\alpha = 0)$, we have
\begin{equation}
\bigl|\eta_\pm(0,t)\bigr|^2 = \omega_0\biggl\{\frac{\cos[\omega_\pm(\lambda')t]^2}{4\omega_\pm(\lambda)} +\frac{\omega_\pm(\lambda)\sin[\omega_\pm(\lambda')t]^2}{4[\omega_\pm(\lambda')]^2}\biggr\}.
\end{equation}
Notice now that in the case in which the initial state is close to critical, the closer $\lambda$ is to $\lambda_c$ the more delocalized the mode $\bar{c}_-$ is initially. This results in the divergence of the amplitude of the oscillations of $\langle \delta X(\alpha)^2\rangle$ as $1/\sqrt{\lambda_c-\lambda}$ when $\lambda\to\lambda_c$. On the other hand, when the final coupling constant $\lambda'$ approaches criticality, one is describing the physics of an initially confined  mode $\bar{c}_-$ that is ``released'' at $t = 0$: it is therefore not surprising that for large times $t$ the width of the distribution increases linearly with time when $\lambda' =\lambda_c$, that is,
\begin{equation}
\langle\delta X(\alpha)^2\rangle  \simeq \frac{\omega_0\omega_-(\lambda)}{4}\,t^2.
\end{equation}

\subsection{Occupation number}

The statistics of the total occupation of the bosonic modes
\begin{equation}
N=a^{\dagger}a+b^{\dagger}b,
\end{equation}
turns out to encode similar information. Let us compute the associated characteristic function
\begin{equation}
G_N(u)=\langle 0_{\lambda} |\, e^{iH[\lambda']t} e^{-iNu} e^{-iH[\lambda']t}\, | 0_{\lambda} \rangle.
\end{equation}
First of all we express the operator $N$ in terms of the modes diagonalizing $H[\lambda']$. Using Eq.~(\ref{eigenmodes}) we obtain $N=N_++N_-$, where
\begin{multline}
N_{\pm}=\cosh^2(\bar{r}_{\pm}) \bar{c}^{\dagger}_{\pm}\bar{c}_{\pm}+\sinh^2(\bar{r}_{\pm}) \bar{c}_{\pm}\bar{c}^{\dagger}_{\pm}\\
-\sinh(\bar{r}_{\pm})\cosh(\bar{r}_{\pm}) [(\bar{c}^{\dagger}_{\pm})^2+(\bar{c}_{\pm})^2)].
\end{multline}
Evolving this operator in time and expressing the initial state $| 0_{\lambda} \rangle$ in terms of the vacuum $| 0_{\lambda'} \rangle$ of the operators $\bar{c}_{\pm}$ as in Eq.~\ref{vacuum}, we obtain $G_{N}(u)=G_{N_+}(u)G_{N_-}(u)$, where
\begin{equation}\label{melem}
G_{N_{\pm}}(u)=e^{iu/2} \left\langle \exp\biggl[\sum_{j=1}^3\gamma_j(\pm)K_j(\pm)\biggr] \right\rangle.
\end{equation}
Here 
\begin{eqnarray}\label{K}
K_1(\pm)&=&K_2^{\dagger}(\pm)=\frac{(\bar{c}_{\pm}^{\dagger})^2}{2},\\
K_3(\pm)&=& \frac{\bar{c}^{\dagger}_{\pm}\bar{c}_{\pm}+\bar{c}_{\pm}\bar{c}^{\dagger}_{\pm}}{2},
\end{eqnarray}
and
\begin{align}
\gamma_1(\pm)&=\gamma_2(\pm)^*=-iu\bigl\{\cosh(2\bar{r}_{\pm})\sinh(2\xi_{\pm})
\nonumber \\ 
&\quad-\sinh(2\bar{r}_{\pm})\cosh(2\xi_{\pm})\cos[2\omega_{\pm}(\lambda')t] \nonumber \\
&\quad-\sinh(2\bar{r}_{\pm})\sin[2\omega_{\pm}(\lambda')t]\bigr\},  \\
\gamma_3(\pm)&= -2iu\bigl\{\cosh(2\bar{r}_{\pm})\cosh(2\xi_{\pm})\nonumber \\ 
&\quad-\sinh(2\bar{r}_{\pm})\sinh(2\xi_{\pm})\cos[2\omega_{\pm}(\lambda')t]
\bigr\}.
\end{align}

In order to compute the matrix elements in Eq.~(\ref{melem}), we notice that for both $+$ and $-$ modes the operators $K_j$ form a closed algebra with commutation relations $[K_1,K_2]=-2K_3$, $[K_1,K_3]=-K_1$, and $[K_2,K_3]=K_2$. We may then apply a standard operator ordering theorem~\cite{Barnett} stating that for this algebra of operators the equality 
\begin{eqnarray}\label{ordering}
e^{\sum_{j=1}^3\gamma_j K_j}=e^{\Gamma_1K_1}e^{\ln(\Gamma_3)K_3}
e^{\Gamma_2K_2},
\end{eqnarray}
holds, where
\begin{align}
\Gamma_{1,2}&=\frac{2\gamma_{1,2}\sinh(\beta)}{2\beta\cosh(\beta)-\gamma_3\sinh(\beta)}, \\
\Gamma_3&=\frac{1}{\biggl[ \cosh(\beta)-\dfrac{\gamma_3}{2\beta}\sinh(\beta) \biggr]^2},
\end{align}
with $\beta^2=\gamma_3^2/3-\gamma_1\gamma_2$. Using Eq.~(\ref{ordering}) together with Eq.~(\ref{K}), we easily
obtain
\begin{eqnarray}
 \left\langle \exp\biggl[\sum_{j=1}^3\gamma_j(\pm)K_j(\pm)\biggr] \right\rangle=(\Gamma_3(\pm))^{\frac{1}{4}}.
\end{eqnarray}
Some straightforward algebra now shows that in the present case $\beta^2=-u^2$ and hence a direct computation of $\Gamma_3$ leads us to
\begin{eqnarray}\label{carN}
G_{N_{\pm}}(u)=\frac{e^{iu/2}}{\sqrt{ \cos(u)+ig_{\pm}(t)\sin(u)}},
\end{eqnarray}
where
\begin{align}
g_{\pm}(t)&=\frac{1}{2}\left[ \frac{\omega_{\pm}(\lambda')^2}{\omega_0\omega_{\pm}(\lambda)}+ \frac{\omega_0\omega_{\pm}(\lambda_0)}{\omega_{\pm}(\lambda')^2} \right]\sin^2[\omega_{\pm}(\lambda')t] \nonumber \\
&\quad+ \frac{1}{2}\left[ \frac{\omega_{\pm}(\lambda) }{\omega_0 } +\frac{\omega_0 }{\omega_{\pm}(\lambda) } \right]\cos^2[\omega_{\pm}(\lambda')t].
\end{align}

The only parameters entering the characteristic function are $g_{\pm}(t)$ ($g_{
\pm}\geq 1$). Physically they characterize the average occupation of the bosonic modes. Indeed, taking the first logarithmic derivative of the characteristic function  leads to
\begin{eqnarray}
\langle N \rangle=\frac{ g_+(t)+g_-(t)}{2}-1.
\end{eqnarray}
As for the average work and the quadrature variance $\langle \delta X(\alpha)^2 \rangle$, the behavior of the occupation for quenches starting close to criticality and going toward criticality is deeply different. When $\lambda' \rightarrow \lambda_c$ we indeed have that for large times
\begin{eqnarray}
\langle N \rangle \simeq \frac{\omega_0\omega_-(\lambda)}{2}\;t^2,
\end{eqnarray}
while in the second case ($\lambda \rightarrow \lambda_c$) the amplitude of the oscillations diverges as $\sqrt{\lambda_c/(\lambda_c-\lambda)}$. 

Finally, we give the result for the full distribution of the occupation number $N$. By taking the Fourier transform of Eq.~\ref{carN}, one may easily obtain
\begin{equation}
P(N)=\sum_{M=0}^{+\infty}\;{\cal N}(M)\; \delta(N-2M),
\end{equation}
where the weights ${\cal N}(M)$ are given by the finite sums
\begin{align}
{\cal N}(M)&=\sqrt{\frac{4}{(1+g_+)(1+g_-)}}\sum_{k=0}^M \left(\begin{matrix} 
      k-\frac{1}{2} \\
      k\\
   \end{matrix}\right)\negthinspace\left(\begin{matrix} 
      M-k-\frac{1}{2} \\
      M-k\\ \end{matrix}\right)\nonumber \\
      &\quad \times \left(\frac{g_+-1}{g_++1}\right)^{2k}  \left(\frac{g_--1}{g_-+1}\right)^{2(M-k)} .
\end{align}

\section{Conclusions}\label{sec3}
In this paper, we studied the statistics of the work and other observables for quantum quenches of a prototypical quantum critical system, the Dicke model. Focusing on quenches of the coupling constant from an initial value $\lambda$ to a final one $\lambda'$, both in the normal phase ($\lambda,\lambda' <\lambda_c$), we computed exactly the characteristic
function of the probability distribution of the work, as well as those associated with the statistics of the quadratures
of the cavity modes and of the total occupation of the bosonic modes. We found that while criticality always leaves an imprint on the dynamics of the system, and particularly on the statistics of observables, there is a deep difference between quenches starting close to the critical point and those going towards it. In the first case the average work (as well as the amplitude of the oscillations of the variance of the quadratures and of the average occupation) diverges as criticality is approached. In contrast, for quenches  towards the quantum critical point the moments of the distribution  of the work stay finite, while the distribution itself displays a sequence of edge singularities. This occurrence is accompanied by a characteristic quadratic growth in time of the variance of the ``coordinate'' operator associated  with the cavity field, and a similar quadratic  temporal growth of the average number of bosonic modes in the system. We developed a simple physical picture explaining the origin of these effects: for quenches starting close to criticality the divergences are caused by a  mode initially
delocalized in phase space subject to a final confining potential that tends to compress it in a finite volume. The situation is opposite for quenches toward criticality: the initially localized mode is released at $t = 0$, and left
to spread coherently in phase space. On the basis of this general qualitative picture, which definitely also holds for
quenches with both $\lambda,\lambda' > \lambda_c$, we expect our main qualitative findings listed above to apply also in this latter case.

\section*{Acknowledgements}
This work is partly based on the Diploma Thesis of F.~P. that was submitted to the Abdus Salam ICTP. A.~S. would like to thank A. Polkovnikov and E. Altman for useful discussions on this and closely related subjects.

\end{document}